\documentstyle[eqsecnum,aps,epsf]{revtex}

\begin{document}
\draft
\title{Chemical potential of the generalized Hubbard model with correlated
hopping}		
\author{V.~Hankevych\cite{e-mail} and L.~Didukh}
\address{Ternopil State Technical University, Department of Physics,\\
56 Rus'ka Str., Ternopil UA-46001, Ukraine }
\date{\today}

\maketitle

\begin{abstract}
In the present paper we study chemical potential of the generalized Hubbard
model with correlated hopping. The peculiarity of the model in comparison with
similar generalized Hubbard models is the concentration dependence of
hopping integrals. 
Chemical potential as a function of the model energy parameters, electron
concentration and temperature is found.
It is shown that correlated hopping and temperature changes essentially the
chemical potential location; these dependencies differ strongly at different
values of the electron concenntration. 
\end{abstract}

\section{Introduction}
One of the simplest models, which is used for a description of strongly
correlated electron systems (in particular, transition metal compounds,
polymers, fullerenes C$_{60}$, high-temperature superconductors, heavy fermion
substances) is the Hubbard model~\cite{hub1} (for reviews see 
papers~\cite{thm}-\cite{geb}). 
Hubbard first showed that taking into account the intra-atomic Coulomb 
repulsion $U$ of two electrons essentially modifies the band energy spectrum
and plays the main role in the formation of electrical and magnetic 
properties of the material with narrow energy bands. 
The ideas of ``lower'' and ``upper'' Hubbard bands, introduced in this paper,
proved to be useful both for our understanding  of physics  of
correlation effects in narrow energy bands and for the interpretation of
experimental data for narrow band materials, in particular high-$T_c$
superconductors. It should be noted that progress in experimental techniques,
such as photoemission and x-ray absorption spectroscopy, has 
recently confirmed and enabled direct measurements of these theoretically 
predicted Hubbard bands~\cite{fuj1}-\cite{bba}.

However, theoretical analyses, on the one hand, and available experimental 
data, which argue an electron-hole asymmetry, on the other hand, point out 
the necessity of the Hubbard model generalization by taking into account
correlated hopping (electron-hole asymmetry is a peculiar property of the
Hubbard model). Really, in some compounds (e.g. see the estimation in 
Refs.~\cite{1_12}-\cite{1_63}) the matrix elements of electron-electron 
interactions describing inter-site hoppings of electrons (correlated 
hopping) are of the same order that the hopping integral or on-site Coulomb 
repulsion; also, the results of the first principles calculations performed by
Hirsch~\cite{hir1} show that hopping integrals of electrons between 
neighbouring ions can depend strongly on the instantaneous charge occupation
of the two atoms involved in the hopping process (in contrast to the Hubbard
model where inter-site hoppings of electrons do not correlate by the 
occupation of these sites). On the other hand, the experimental observations
show electron-hole asymmetry of properties of strongly correlated electron
systems. In particular, transition metal oxides exhibit the electron-hole 
asymmetry of conductivity~\cite{jon}, transition $3d$-metals have the 
electron-hole asymmetry of cohesive energy~\cite{say,rkh}. In this connexion,
the experimentally observed~\cite{fug,kan} electron-hole asymmetry of 
superconducting properties of high-temperature superconductors should be 
also noted.

Consequently, in recent years the generalized Hubbard model with 
correlated hopping has been used widely to describe strongly correlated 
electron systems~\cite{sv1}-\cite{1_51}; the electron-hole asymmetry
is a property of such a generalized Hubbard model as a result of the 
dependence of the hopping integral on the occupation of the sites involved 
in the hopping process. 

Generalized Hubbard model has been used for an investigation of metallic
ferromagnetism in narrow energy bands~\cite{dps,sv1}-\cite{ah}. In 
particular, a generalization of Nagaoka's theorem has been proved in
paper~\cite{ksv}, and it has been shown~\cite{dps,ah} that in strong coupling regime
the model strongly favours ferromagnetism for electron concentration $n>1$ 
versus electron concentration $n<1$, as observed in transition metal 
compounds.

Electrical properties, metal-insulator transition and high-temperature
superconductivity have been studied in the generalized Hubbard model
in papers~\cite{1_63,1_6}-\cite{1_51}. The result of these papers is
the generalized Hubbard model has much better physics 
than the Hubbard model, firstly, and usage of the electron-hole asymmetry 
concept allows one to interpret the peculiarities of physical properties of 
strongly correlated electron systems, which are not explained by the Hubbard 
model, secondly. In particular, the experimentally observed 
electron-hole asymmetries of metal oxides conductivity, of cohesive energy of transition 
$3d$-metals and of superconducting properties of high-temperature 
superconductors, mentioned above, have been explained within the generalized 
Hubbard model with correlated hopping in 
papers~\cite{1_6,1_26,dhs}, \cite{1_6}, \cite{hir2} respectively.

An important puzzle remaining beyond consideration in the cited papers
is the question about location of the model chemical potential and
its dependence on electron concentration and temperature. Theoretical 
calculations of chemical potential dependences can be compared with the 
experimental data obtained by photoemission spectroscopy methods. Also these 
dependences of chemical potential could give indication for the existence of 
phase separation in system upon doping.

Even in the framework of the Hubbard
model this problem is still open, in particular the question about
the jump in chemical potential of the Hubbard model at half-filling 
in going from electron to hole doping is under 
discussion in recent literature (see, for instance, review~\cite{gkk}).
On the one hand, in majority of papers it has been found that the jump
in chemical potential at half-filling is equal to width of the energy gap.
On the other hand, some authors suggest that the discontinuity in chemical
potential is strictly smaller than the energy gap as a result of inducing 
states inside the Mott-Hubbard gap (so called ``midgap states'') upon doping
the Mott insulators.
The first scenario is confirmed by the exact solution~\cite{4_5} of the 
Hubbard model in one dimension. Also numerical studies of the two-dimensional 
Hubbard model performed in Ref.~\cite{4_6} by means of the Monte Carlo 
methods show that when the electron
concentration $n=1-\delta$ ($\delta$ is infinitesimal doping), the chemical
potential is located at the top of the lower Hubbard band; when the electron
concentration $n=1+\delta$, the chemical potential jumps to the bottom of
the upper Hubbard band. Correspondingly, upon electron doping the chemical 
potential jumps from the top of the lower Hubbard band to the bottom of the 
upper Hubbard band according to the first scenario. The results of 
papers~\cite{4_8,4_9} obtained by means of the Kotliar-Ruckenstein slave
bosons~\cite{1_53} for the three dimensional Hubbard model and the quantum 
Monte Carlo
method in infinite dimensions~\cite{3_12} suggest that the jump
in chemical potential at half-filling is equal to width of the energy gap.

Using the projective self-consistent technique~\cite{4_4c} in infinite 
dimensions, the authors of papers~~\cite{4_11} found that, for any value
of $U$ larger than $U_c$, doping does induce states inside the Mott-Hubbard
gap; the splitting of these states from the edge of the Hubbard band is 
a fraction of the bare kinetic energy, and remains finite even in the limit 
where $U$ is infinite.
Therefore the jump in chemical potential for infinitesimal doping is 
strictly less
than the energy gap, so the second scenario described above holds.
These conclusions were also supported by an extention~\cite{4_12,1_71} of 
the iterated perturbative theory approximation~\cite{gkk} for the Hubbard 
model away from half-filling. Midgap states have also been 
predicted~\cite{4_13} to occur upon doping in the infinite dimensional
Hubbard model using an analytic variant of the Lanczos continued fraction
method~\cite{hk}.

In consequence of the absence of a rigorous result on the described 
scenarios adequacy the subject is still open. Moreover, even in infinite 
dimensions this problem is under discussion: different approximations,
for example the quantum Monte Carlo method and the iterated perturbative 
theory, lead to the contrary results. Another question is how correlated
hopping does influence on chemical potential. 
Therefore, the investigation of 
chemical potential behaviour in the doped Hubbard model as well as in the 
generalized Hubbard model with correlated hopping is a relevant and 
important task, which the present paper is devoted.

The paper has the following structure. In Section~2 we formulate Hamiltonian 
of the generalized Hubbard model with correlated hopping, the quasiparticle 
energy spectrum of the model is derived by means of a generalized mean-field 
approximation (an analogue of the projection operation) in the Green function 
technique. The general properties of the spectrum, and the approximation
used by us are discussed. In Sections~3-4 the dependences of chemical potential
on the model energy parameters, correlated hopping, temperature and electron
concentration are studied. The obtained results are compared with the 
experimental data on Fermi energy behaviour of real strongly correlated
electron systems. Finally, Section~5 is devoted to the conclusions from the 
obtained results.

\section{Hamiltonian and quasiparticle energy spectrum of the model}

We start from the natural generalization of the Hubbard model including
the matrix elements of electron-electron interaction~\cite{1_12,dps}
\begin{eqnarray}
J(ikjk)=\int \int \varphi^*({\bf r}-{\bf R}_i)\varphi({\bf r}-{\bf R}_j)
{e^2\over |{\bf r}-{\bf r'}|}|\varphi({\bf r'}-{\bf R}_k)|^2{\bf drdr'},
\end{eqnarray}
which correlate the electron hopping from one lattice site to another
($\varphi({\bf r}-{\bf R}_i)$ is the Wannier function). 
The Hamiltonian of such a generalized Hubbard model with correlated
hopping reads as~\cite{1_6,1_62}
\begin{eqnarray} \label{ham}
&&H=H_0+H_1+H'_1, \\
&&H_0=-\mu \sum_i \left(X_i^{\uparrow}+X_i^{\downarrow}+2X_i^2\right)+
U\sum_{i}X_i^2,\\
&&H_1=t(n){\sum \limits_{ij\sigma}}'X_i^{\sigma 0} X_j^{0\sigma} +
\tilde{t}(n){\sum \limits_{ij\sigma}}'X_i^{2\sigma}X_j^{\sigma 2},
\\
&&H'_1=t'(n){\sum \limits_{ij\sigma}}' \left(\eta_{\sigma}X_i^{\sigma 0}
X_j^{\bar{\sigma} 2}+h.c.\right),
\end{eqnarray}
where $\mu$ is the chemical potential, the primes at the sums signify 
that $i\neq j$, $X_i^{kl}=|k\rangle\langle l|$ is
the Hubbard operator~\cite{hubb2}, $X_i^k=X_i^{kl}X_i^{lk}$ is the operator 
of the number of $|k\rangle$-states on $i$-site; $|0\rangle$ denotes the 
state of site, which is not occupied by an electron (hole), 
$|\sigma \rangle$ denotes the state of singly occupied (by an electron with 
spin $\sigma$) $i$-site, $|2\rangle$ denotes the state of doubly occupied 
(by two electrons with the opposite spins) $i$-site (doublon), 
$\eta_{\uparrow}=-1,\ \eta_{\downarrow}=1$.
\begin{eqnarray} \label{c_d_h}
&&t(n)=t_0+n\sum_{\stackrel{k\neq{i}}{k\neq{j}}}J(ikjk)=t_0+nT_1,
\\
&&\tilde{t}(n)=t(n)+2T_2, \quad
t'(n)=t(n)+T_2, \quad T_2=J(iiji)=J(iiij),
\end{eqnarray}
where $t_0$ is the uncorrelated hopping integral (the matrix element of 
electron-ion interaction), $n=N_e/N$ is the electron concentration
($N_e$ is the number of electrons within conduction band, 
$N$ is the number of lattice sites).

$H_0$ describes the atomic limit of narrow-band models.
$H_1$ describes the translational hopping of holes and doublons;
within the present model the hopping integrals of holes $t(n)$ and 
doublons $\tilde{t}(n)$ are different (in contrast with narrow-band models of 
the Hubbard type), this leads to the electron-hole asymmetry mentioned above.
$H'_1$ describes the processes of paired creation and destruction of holes
and doublons.

In the model described by Hamiltonian~(\ref{ham}) an electron hopping from 
one site to another is correlated both by the occupation of the sites 
involved in the hopping process (with the hopping integral $T_2$) and the 
occupation of the nearest-neighbour sites (with the hopping integral $T_1$)
which we took into account by means of the Hartree-Fock approximation
(see expression~(\ref{c_d_h})).
The peculiarity of model~(\ref{ham}) is taking into consideration of the
correlated hopping $T_1$ which cause the concentration dependence 
of the hopping integrals in contrast with similar generalized Hubbard models. 

For half-filling case we introduce the following notations: 
$t(n)\equiv t=t_0+T_1,\ \tilde{t}(n)\equiv\tilde{t}=t+2T_2,\ 
t'(n)\equiv t'=t+T_2$. In this case for $t'=0$ some exact results have
been found~\cite{sv1,1_16,ovch,bs}. In a simple cubic
lattice with coordination number $z$ metal-insulator transition occurs at
\begin{eqnarray} \label{exact_cr}
U_c=2z|t_0|.
\end{eqnarray}   
If $U>U_c$ the ground state of system is a paramagnetic Mott-Hubbard 
insulator with the concentration of doubly occupied sites $d=0$, the ground
state energy is equal to zero.

The single-particle Green function in terms of the Hubbard operators reads
as
\begin{eqnarray} \label{spgf}
\langle\langle a_{p\sigma}\!|\hfill a_{p'\sigma}^{+}
\rangle\rangle
=\langle\langle
X_p^{0\sigma}|X_{p'}^{\sigma 0}\rangle\rangle +\eta_{\sigma}\langle\langle
X_p^{0\sigma}|X_{p'}^{2\bar{\sigma}}\rangle\rangle +\eta_{\sigma}\langle\langle 
X_p^{\bar{\sigma} 2}|X_{p'}^{\sigma 0}\rangle\rangle 
+\langle\langle 
X_p^{\bar{\sigma} 2}|X_{p'}^{2\bar{\sigma}}\rangle\rangle.
\end{eqnarray}
The Green function 
$\langle\langle X_p^{0\sigma}|X_{p'}^{\sigma 0}\rangle\rangle$ 
is given by the equation
\begin{eqnarray} \label{eq}
(E+\mu)\langle\langle X_p^{0\sigma}|X_{p'}^{\sigma 0}\rangle\rangle=&&
{\delta_{pp'}\over 2\pi}\langle X_p^{\sigma}+X_p^{0}\rangle+ 
\langle\langle\left[X_p^{0\sigma}, H_1\right]|X_{p'}^{\sigma 0}\rangle\rangle
\nonumber\\
&&+\langle\langle\left[X_p^{0\sigma}, H'_1\right]|X_{p'}^{\sigma 0}\rangle
\rangle,
\end{eqnarray} 
with $[A, B]=AB-BA$. Using a variant~\cite{1_6,1_61} of the generalized
mean-field approximation~\cite{appr} we suppose in Eq.~(\ref{eq}) that
\begin{eqnarray}
\left[X_p^{0\sigma}, H_1\right]\simeq\sum_{j}\epsilon(pj)X_j^{0\sigma},
\quad
\left[X_p^{0\sigma}, H'_1\right]\simeq\sum_{j}\epsilon_1(pj)X_j^{\bar{\sigma}2},
\end{eqnarray}
where $\epsilon(pj)$ and $\epsilon_1(pj)$ are the non-operator expressions. 
The procedure of $\epsilon(pj)$ and $\epsilon_1(pj)$ calculation
is described in Ref.~\cite{1_6,1_62} (here there is a partial equivalence 
with the slave boson method~\cite{1_53}). 

Thus we obtain the closed system of equations for the Green functions 
$\langle\langle X_{p}^{0\sigma}| X_{p'}^{\sigma 0}\rangle\rangle$ and
$\langle\langle X_{p}^{\bar{\sigma}2}| X_{p'}^{\sigma 0}\rangle\rangle$. 
An analogous procedure is realized also in the equations for the other
Green functions~(\ref{spgf}).

In this way, we find single-particle Green function~(\ref{spgf}) and 
quasiparticle energy spectrum. For paramagnetic case in 
${\bf k}$-representation the spectrum is~\cite{1_26}:
\begin{eqnarray} \label{es}
&&E_{1,2}({\bf k})=-\mu+{U\over 2}+{\epsilon({\bf k})+\tilde{\epsilon}
({\bf k})\over 2}\mp {1\over 2}Q({\bf k}),\\
&&Q({\bf k})=\sqrt{[\epsilon({\bf k})-\tilde{\epsilon}({\bf k})-U]^2+
4\epsilon_1({\bf k})\epsilon_2({\bf k})}. \label{q}
\end{eqnarray} 
The Fourier components of the quantities defining quasiparticle energy
spectrum~(\ref{es}) are given by the formulae: 
\begin{eqnarray} \label{ep}
\epsilon({\bf k})=\alpha t_{\bf k}(n),\quad 
\tilde{\epsilon}({\bf k})=\tilde{\alpha} \tilde{t}_{\bf k}(n),\quad  
\epsilon_1({\bf k})=\alpha_1 t'_{\bf k}(n),\quad
\epsilon_2({\bf k})=\alpha_2 t'_{\bf k}(n),
\end{eqnarray}
\begin{eqnarray}
&&\alpha=-n+2d+{2(1-d)^2\over 2-n}-{2d(d-n+1)\over 2-n}{\tilde{t}(n)
\over t(n)},\\
&&\tilde{\alpha}=n-2d+{2d^2\over n}-{2d(d-n+1)\over n}{t(n)\over \tilde{t}(n)},
\\
&&\alpha_1=n-1-{2d\over n}, \quad 
\alpha_2=-1-n+{2(1-d)\over 2-n},
\end{eqnarray}
with $d$ beeing the concentration of doubly occupied sites.

For $n<1$ and $U\to \infty$ we obtain
\begin{eqnarray} \label{es<}
E_1({\bf k})=-\mu+\left({2\over 2-n}-n\right)t_{\bf k}(n)
\end{eqnarray}
(the lower Hubbard band); if $n>1$ and only the upper Hubbard band is 
important, we obtain
\begin{eqnarray} \label{es>}
E_2({\bf k})=-\mu +U+\left({2\over n}-2+n\right)\tilde{t}_{\bf k}(n).
\end{eqnarray}
For the Hubbard model ($t(n)=\tilde{t}(n)=t'(n)=t_0$) $E_1({\bf k})$ for
$n\to 0$, and $E_2({\bf k})$ for $n\to 2$ get the band form. Moreover,
expressions~(\ref{es}) describe the exact atomic limit for $t_0=0$ and
the band situation for $U=0$.

At half-filling qausiparticle energy spectrum~(\ref{es}) is writen in 
the form~\cite{1_62}:
\begin{eqnarray} \label{es1}
&&E_{1,2}({\bf k})=-\mu+{(1-2d)(t_{\bf k}+\tilde{t}_{\bf k})+U\over 2}\mp 
{1\over 2}F_{\bf k},
\\
&&F_{\bf k}=\sqrt{\left[B(t_{\bf k}-\tilde{t}_{\bf k})-U\right]^2+
(4dt'_{\bf k})^2},\
B=1-2d+4d^2,
\end{eqnarray}
where the doublon concentration $d$ is found from the equation
\begin{eqnarray} \label{d}
d=\langle X_i^2\rangle ={1\over 2N}\sum_{\bf k}\left(
{A_{\bf k}\over \exp{E_1({\bf k})\over \theta}+1}+
{B_{\bf k}\over \exp{E_2({\bf k})\over \theta}+1}\right),
\end{eqnarray} 
with
\begin{eqnarray}
&&A_{\bf k}={1\over 2}-{B(\tilde{t}_{\bf k}-t_{\bf k})\over 2F_{\bf k}}-
{U\over 2F_{\bf k}}, \qquad
B_{\bf k}={1\over 2}+{B(\tilde{t}_{\bf k}-t_{\bf k})\over 2F_{\bf k}}+
{U\over 2F_{\bf k}},
\end{eqnarray}
$\theta=k_BT$, $k_B$ is the Boltzmann's constant, $T$ is the temperature.
For the special case $t'=0$ of the model quasiparticle energy 
spectrum~(\ref{es1}) reproduces (see Ref.~\cite{1_62}) exact 
result~(\ref{exact_cr}).

As a result of the absence of a natural small expansion parameter 
at intermediate to strong Coulomb interactions in the Hubbard model and its 
generalizations (this situation is under consideration in the present paper), 
one does not know a rigorously justified nonperturbative 
approach in the theory of strongly correlated electron systems. In such a 
case the methods of mean-field type are useful~\cite{gkk,geb}, we have 
used one of these methods. The justification of an approximation
is the reproduction of limiting cases and known
exact results, and also the accordance with experimental data.
Discussing the generalized mean-field approximation used by us we note the 
following facts.

The approximation reproduces the exact atomic and band limits in the Hubbard
model and also describes metal-insulator transition as mentioned above. 
For the Hubbard model the criterion of metal-insulator transition obtained 
by means of this approximation is $U_c=2w_0$ ($w_0=z|t_0|$) in agreement 
with the Mott's criterion~\cite{mott} and the result of the random dispersion 
approximation~\cite{geb} in infinite dimensions. 
In the limit of strong interactions the approach satisfies another 
requirement of a good approximate theory for the Hubbard model at 
half-filling formulated in Ref.~\cite{geb}: $W_{l,u}=2w_0$ ($W_l,\ W_u$ are
the widths of the lower and upper Hubbard bands). For the special case 
$t'=0$ of the model~(\ref{ham}) the generalized mean-field approximation 
reproduces exact criterion~(\ref{exact_cr}) of metal-insulator
transition, the ground state energy and doublon concentration.
For very small (large) values of electron concentration and $U\to \infty$
the approximation gives the band description according to the general 
physical ideas. The approach allows~\cite{1_6} also to describe the observed 
transition in 
some Mott-Hubbard compounds from a metallic state to an insulating phase with 
the increase of temperature in a paramagnetic state (see, for example, 
book~\cite{mott}, p.~178, 195). 

On the other hand, a deficiency of the 
generalized mean-field approximation is the breakdown of
Fermi liquid behaviour of the systems with weak interactions and half-filled 
band (however, it should be noted that in the present paper we consider
the cases of strong and intermediate interactions). The question about
Fermi liquid behaviour (i.e. the fulfilment of Luttinger's theorem~\cite{lut})
at intermediate to strong interactions in three dimensions is under
discussion~\cite{gkk,geb}. This is caused by the reason that Luttinger's
theorem is based~\cite{lut,pwa} on the perturbation expansion, assuming an 
adiabatic relation between a noninteracting system and an interacting one, 
which may not be fulfiled for the case of intermediate and strong 
interactions. In particular, the authors~\cite{pls,gze} have found the 
violation of Luttinger's theorem in the doped Hubbard model, and $t$-$J$
model which is a partial case of the former one. Therefore, although
the system is metallic away from half-filling, it may be non-Fermi
liquid at intermediate and strong interactions because of a violation of 
Luttinger's theorem.

Another limitation of
the generalized mean-field theory is that for small values $t'/U$ the 
approximaton does not reproduce the physics of the interaction of local 
magnetic moments as this holds in the framework of the effective Hamiltonian 
method. Nevertheless, we consider the paramagnetic phase, and in this case
the effective inter-atomic exchange interaction (kinematic superexchange)
leads to the renormalization of chemical potential only (in the molecular field
approximation). Thus, apparently, it can be assumed that the generalized
mean-field approximation used by us can provide a good description of
effects of electron correlations in the paramagnetic phase and the region of 
intermediate (when a metal-insulator transition can occur) and strong 
interactions. Note also, that a modification~\cite{ld_na} of this approximation 
proposed one of us allows to take into account effects caused by
the effective interaction of local magnetic moments; at the same time, the
main features of quasiparticle energy spectrum~(\ref{es}) still persist.

The noted generalized mean-field approximation has been applied also for 
study of 
metal-insulator transition and properties of generalized Hubbard model with 
correlated hopping~\cite{1_61,1_62,1_26,dhs}, for description of 
metal-insulator
transition in the doubly degenerate Hubbard model~\cite{1_61,dsdh1} and 
a doubly orbitally degenerate model with correlated hopping~\cite{dsdh2}.

\section{Chemical potential of the model at half-filling}
\setcounter{equation}{0}

Consider the important situation of half-filled band $n=1$.
Chemical potential of the model is given by the equation 
($\langle X_i^0\rangle=\langle X_i^2\rangle$):
\begin{eqnarray} \label{ch_p}
&&\sum_{\bf k}\left({A_{\bf k}\over \exp{E_1({\bf k})\over \theta}+1}+
{B_{\bf k}\over \exp{E_2({\bf k})\over \theta}+1}\right)
\nonumber\\
&&= \sum_{\bf k}\left({B_{\bf k}\over \exp{-E_1({\bf k})\over \theta}+1}+
{A_{\bf k}\over \exp{-E_2({\bf k})\over \theta}+1}\right).
\end{eqnarray}
From Eq.~(\ref{ch_p}) for the rectangular density of states and $T=0$ we find
chemical potential of the generalized Hubbard model with correlated hopping 
in the region of metal-insulator transition:
\begin{eqnarray} \label{ch_p_m}
&&\mu={w\over w+\tilde{w}}U \qquad (U\le w+\tilde{w}), \\
&&\mu ={U\over 2}+{w-\tilde{w}\over 2} \qquad (U>w+\tilde{w}), \label{ch_p_d}
\end{eqnarray}
with  $w=z|t|,\ \tilde{w}=z|\tilde{t}|$ beeing the half-widths of the lower 
and upper Hubbard band respectively.
For the Hubbard model formulae~(\ref{ch_p_m}) and (\ref{ch_p_d}) give
the well-known result $\mu=U/2$. At $t'=0$ chemical potential of the
generalized Hubbard model is equal to $\mu=U/2$ beeing a consequence of 
the electron-hole symmetry which is a characteristic 
of the model in this case. Note that the value $U_c=w+\tilde{w}$ 
corresponds to the metal-insulator transition point of the generalized 
Hubbard model with correlated hopping.

At arbitrary value of temperature from Eq.~(\ref{ch_p}) we find the 
expression for calculating chemical potential:
\begin{eqnarray} \label{ch_p_T}
\int\limits_{-w}^{w}\left[{1\over \exp{-E_2(\epsilon)
\over \theta}+1}-{1\over \exp{E_1(\epsilon)\over \theta}+1}
\right]d\epsilon=0,
\end{eqnarray}
where $E_1(\epsilon)$, $E_2(\epsilon)$ are obtained from the respective
formulae~(\ref{es1}) for $E_1({\bf k})$ and $E_2({\bf k})$ 
substituting $t_{\bf k}\!\rightarrow\! \epsilon$, 
$\tilde{t}_{\bf k}\!\rightarrow\!{\tilde{t}\over t}\epsilon$, 
$t'_{\bf k}\!\rightarrow\!{t'\over t}\epsilon$.

\begin{figure}[hpb] 
\begin{minipage}[b]{75mm}
\epsfxsize=75mm
\epsfysize=65mm
\epsfclipon
\epsffile{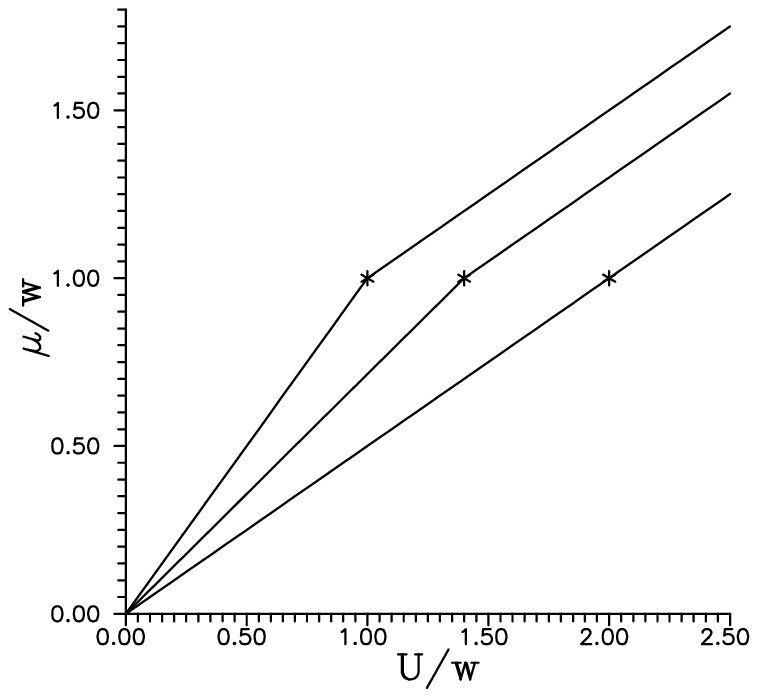}
\renewcommand{\baselinestretch}{1}
\caption{Dependence of chemical potential $\mu$ of the generalized Hubbard 
model with correlated hopping in the ground state: the upper curve 
corresponds to $\tau_1=0,\ \tau_2=0.5$; the middle curve -- 
$\tau_1=0,\ \tau_2=0.3$; the lower curve -- $\tau_1=\tau_2=0$ (the 
Hubbard model). The asterisks denote the point of metal-insulator 
transition.}  \label{0mu_U}
\end{minipage}
\hfill
\begin{minipage}[b]{75mm}
\epsfxsize=75mm
\epsfysize=65mm
\epsfclipon
\epsffile{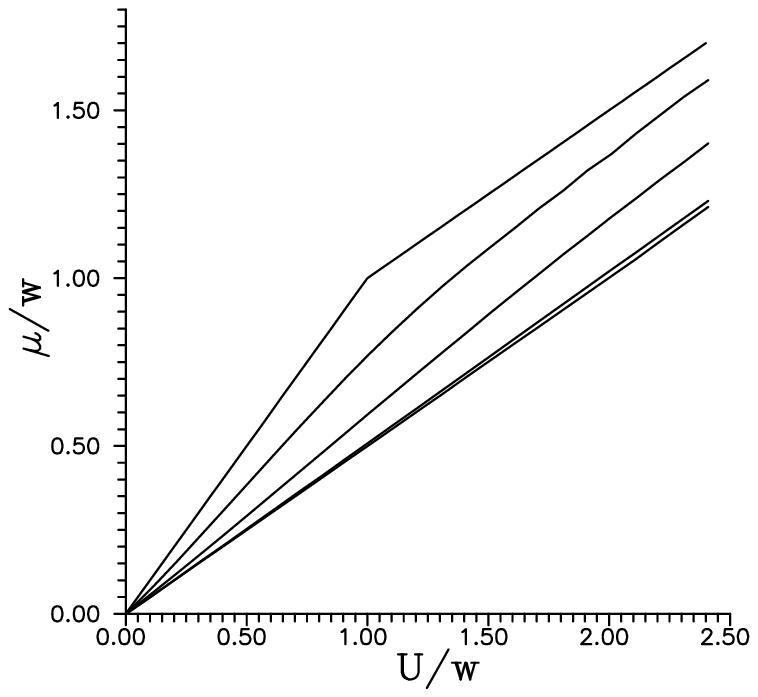}
\renewcommand{\baselinestretch}{1}
\caption{Evolution of the dependence of chemical potential $\mu$ of 
the generalized Hubbard model ($\tau_1=0,\ \tau_2=0.5$) on $U/w$ with 
temperature $\theta/w=k_BT/w=0,\ 0.1,\ 0.3,\ 1$ respectively. The lowermost curve  
corresponds to the behaviour of chemical potential of the Hubbard 
model.} \label{Tmu_U}
\end{minipage}
\vfill	
\begin{minipage}[t]{75mm}
\epsfxsize=75mm
\epsfysize=65mm
\epsfclipon
\epsffile{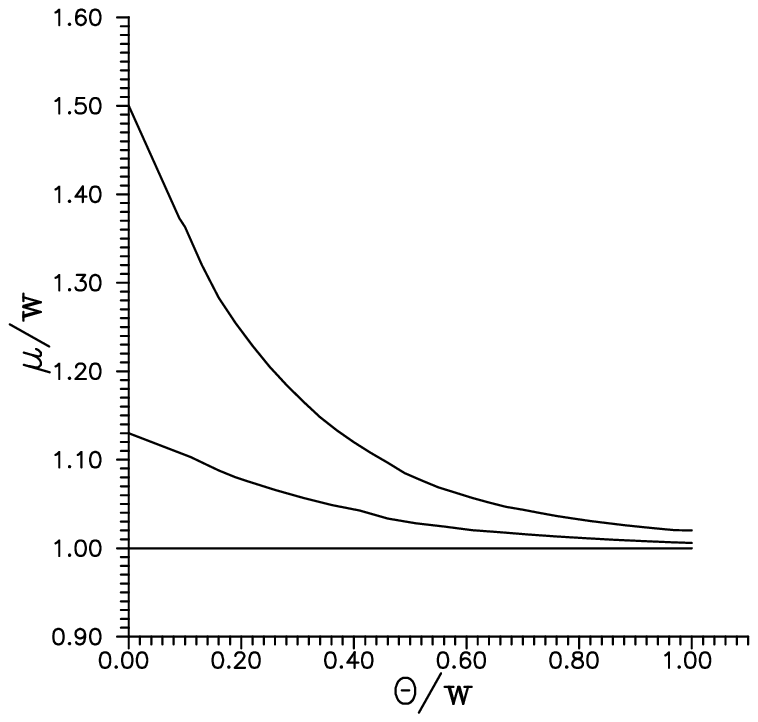}
\renewcommand{\baselinestretch}{1}
\caption{Temperature dependence of chemical potential $\mu$ of 
the generalized Hubbard model with correlated hopping at $U/2w=1$: the upper
curve corresponds to $\tau_1=\tau_2=0.3$; 
the lower curve -- $\tau_1=\tau_2=0.1$; 
the straight -- $\tau_1=\tau_2=0$ (the Hubbard model).}  \label{mu_T}
\end{minipage}
\hfill
\begin{minipage}[t]{75mm}
\epsfxsize=75mm
\epsfysize=65mm
\epsfclipon
\epsffile{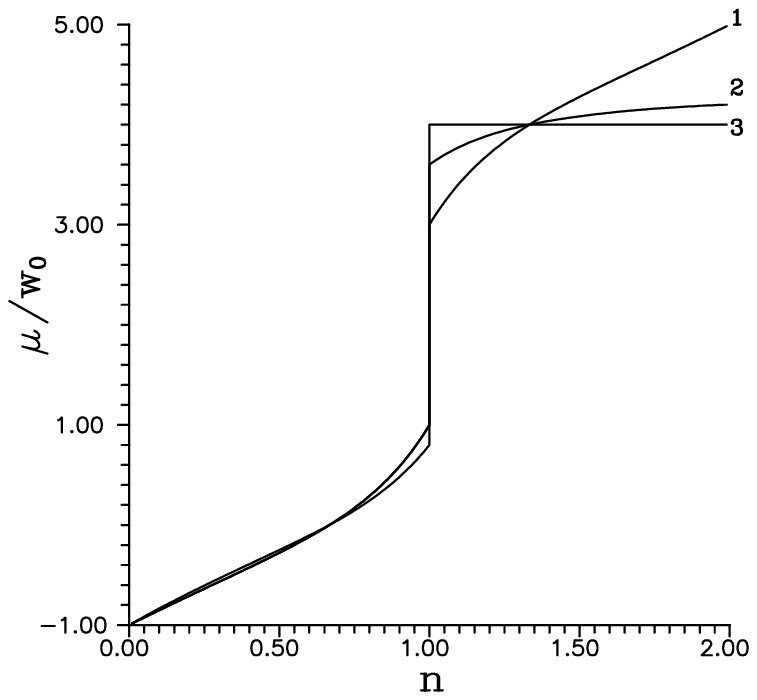}
\renewcommand{\baselinestretch}{1}
\caption{Chemical potential of the generalized Hubbard model versus the
electron concentration $n$ at $U/w_0=4$: $\tau_1=\tau_2=0$ (1); 
$\tau_1=\tau_2=0.2$ (2); $\tau_1=0,\ \tau_2=0.5$ (3).} \label{0mu_n}
\end{minipage}
\end{figure}

Figs.~\ref{0mu_U}-\ref{mu_T}, where the dependences of chemical potential 
on the ratio $U/w$ and temperature are plotted, show that chemical potential 
of the generalized Hubbard model with correlated hopping $\mu >U/2$ and only 
in the absence of correlated hopping or within high temperature region 
the chemical potential is $\mu =U/2$. From Eqs.~(\ref{ch_p_m}) and 
(\ref{ch_p_d}) one can see that chemical potential of the generalized Hubbard 
model with correlated hopping becomes dependent on $U,\ w,\ \tilde{w}$. 
The dependences on these parameters are 
different in metallic and insulating phases, this leads to a kink at the
point of metal-insulator transition (see Fig.~\ref{0mu_U}); 
$\tilde{t}=0$ (i.e. $\tilde{w}=0$) and $t'=0.5t$ correspond to the values of 
correlated hopping parameters $\tau_1=T_1/|t_0|=0$ and $\tau_2=T_2/|t_0|=0.5$. 
Taking into consideration correlated hopping leads to the non-equivalence of 
the lower and upper Hubbard bands. This changes the value of chemical 
potential from the point $\mu=U/2$: with the increase of the correlated 
hopping $T_2$ (i.e. with decreasing $\tilde{w}$) chemical potential 
moves towards the upper Hubbard band. 

With the increase of temperature the kink in the chemical potential curve
disappears (Fig.~\ref{Tmu_U}). Similar peculiarity in the evolution of free 
energy dependence on parameter $U/w$ with temperature
was noted by Mott~\cite{mott}. 

From Fig.~\ref{mu_T} one can see that in the model under consideration 
within low and room temperature regions chemical potential is essentially 
dependent not only on the parameters $w$ and $\tilde{w}$, but becomes also 
dependent on temperature (in contrast to chemical potential of the 
Hubbard model). Moreover, with the decrease of temperature chemical 
potential rapidly increases, this dependence is different for the different
values of correlated hopping parameters $\tau_1,\ \tau_2$. The found 
temperature dependence of chemical potential can be explained by the 
following reasons. At zero temperature in an insulating phase 
the chemical potential is located at the centre of the energy gap 
(between the top of the lower Hubbard band and the bottom of the upper 
Hubbard band), the chemical potential value is larger than $U/2$. 
As the temperature increases some electrons are thermally 
activated and the energy levels within the upper (lower) Hubbard band become
to be occupied (empty). The width of the lower Hubbard band is larger than
the width of the upper Hubbard band, and thus, the density of states in
the lower band is smaller than the density of states in the upper band
(as a result that the distance between the nearest energy levels within the 
lower Hubbard band is larger versus the corresponding distance in the upper 
Hubbard band). Consequently, with increasing temperature chemical potential 
position moves towards the lower Hubbard band, and chemical potential value 
decreases.

Within high temperature region in the generalized Hubbard model chemical
potential tends to $U/2$ with the increase of temperature; really, for
$T\to \infty$ the probabilities of an electron finding within the lower and
upper Hubbard bands are equal. 

\section{Chemical potential of the doped Mott-Hubbard systems}
\setcounter{equation}{0}

Find the dependence of chemical potential $\mu$ of the generalized Hubbard
model with correlated hopping on carrier concentration. Chemical potential is 
given by the equation ($\langle X^2_i\rangle-\langle X^0_i\rangle=n-1$):
\begin{eqnarray} \label{ch_p_eq}
&&{n\over 2N}\sum_{\bf k}\left({C_{\bf k}\over \exp{E_1({\bf k})\over \theta}+1}+
{D_{\bf k}\over \exp{E_2({\bf k})\over \theta}+1}\right)
\nonumber\\
&&-{1-n/2\over N} 
\sum_{\bf k}\left({D_{\bf k}\over \exp{-E_1({\bf k})\over \theta}+1}+
{C_{\bf k}\over \exp{-E_2({\bf k})\over \theta}+1}\right)=n-1,
\end{eqnarray}
with $E_1({\bf k})$ and $E_2({\bf k})$ determining by expression~(\ref{es}),
\begin{eqnarray} \label{4_1_0}
&&C_{\bf k}={1\over 2}-{\tilde{\epsilon}({\bf k})-\epsilon({\bf k})
+U\over 2Q({\bf k})}, \qquad
D_{\bf k}={1\over 2}+{\tilde{\epsilon}({\bf k})-\epsilon({\bf k})
+U\over 2Q({\bf k})}, 
\end{eqnarray}
$Q({\bf k})$ is defined by formula~(\ref{q}), and 
$\tilde{\epsilon}({\bf k}),\ \epsilon({\bf k})$ are given by 
expressions~(\ref{ep}).

Eq.~(\ref{ch_p_eq}) at $T=0$ is rewritten in the form:
\begin{eqnarray} \label{4_1_2}
&&{1\over N}\sum_{\bf k}\left[ D_{\bf k}(n-1)+1-n/2\right]
\theta (E_2({\bf k}))
\nonumber\\
&&+{1\over N}\sum_{\bf k}\left[ D_{\bf k}(1-n)+1/2\right]
\theta (E_1({\bf k}))= 1-n/2.
\end{eqnarray}
We assume the rectangular density of states and pass in Eq.~(\ref{4_1_2})
from a summation over ${\bf k}$ to an integration over energy.
We solve this equation for the case of a paramagnetic phase of the doped 
Mott-Hubbard insulator (the energy spectrum is supposed to has 
form~(\ref{es<}), (\ref{es>})), and find that chemical potential of the 
generalized Hubbard model is
\begin{eqnarray} \label{mu<}
&&\mu ={(3n-2)(2-2n+n^2)\over (2-n)^2}w(n) \qquad (n<1),
\\
&&\mu =U+{(3n-4)(2-2n+n^2)\over n^2}\tilde{w}(n) \qquad (n>1), \label{mu>}
\end{eqnarray}
with $w(n)=w_0(1-n\tau_1),\ \tilde{w}(n)=w_0(1-n\tau_1-2\tau_2)$.

Concentration dependence of the chemical potential~(\ref{mu<}) and (\ref{mu>})
is plotted in Fig.~\ref{0mu_n}. For the Hubbard model this dependence
is in qualitative agreement with the results of Refs.~\cite{4_4a,4_4b},
obtained by means of the Gutzwiller variational method~\cite{1_7} in three 
dimensions
and the projective self-consistent technique~\cite{4_4c} in infinite 
dimensions. However, the curve camber in Fig.~\ref{0mu_n} is as opposed to
that plotted in paper~\cite{4_4a}: the second derivatives of
chemical potential with electron concentration $n$ are opposite in sign;
thus, the behaviour of electron compressibility of system (which is 
defined by the quantity $d\mu\over dn$), determining with the help of 
Eq.~(\ref{mu<}) and (\ref{mu>}), is inverse to the compressibility behaviour
obtained using $\mu (n)$-dependence of Ref.~\cite{4_4a}. The results
of works~\cite{1_71,4_4b} coincide with the conclusions derived from
Fig.~\ref{0mu_n}, however, there is another essential difference between 
them (see the next paragraph). It should be also noted that our calculations
show ${d\mu\over dn}>0$ in all region of the electron concentration. 
Thus we do not predict the existance of phase separation upon doping in the 
framework
of the model under consideration in agreement with the quantum Monte-Carlo
results~\cite{fj} for the Hubbard model in infinite dimensions. Such an instability is 
experimentally observed in high-temperature superconductors~\cite{eme}
and manganites with colossal magnetoresistance~\cite{lok}. This 
interesting feature was predicted theoretically to be a most common behaviour 
of extended Hubbard model with nearest-neighbour Coulomb 
repulsion and double exchange model~\cite{khom}, $t$-$J$ model~\cite{eme}, 
and pseudospin-electron model~\cite{stas}.

Fig.~\ref{0mu_n} displays the jump in chemical potential at half-filling 
in going from electron to hole doping, this discontinuity in chemical 
potential is strictly equal to the energy gap width $\Delta E=U-w-\tilde{w}$.
Therefore, our calculations confirm the first scenario of the chemical
potential evolution upon doping mentioned in Section~1. As a result
of the absence of a rigorous result on the chemical potential behaviour,
one of the main tests of an approximate theory is the comparison with 
experimental data. One knows no experimental data, which show that the 
theoretical prediction of some authors, such as the existence of mid-gap 
states in the Mott-Hubbard insulator upon doping, indeed occur in real
transition metal compounds. However, the experiment~\cite{4_15,ift}
suggests the jump in Fermi energy near half-filling is equal to the 
Mott-Hubbard gap in some strongly correlated electron systems.

Fig.~\ref{0mu_n} shows that correlated hopping changes essentially the 
position of chemical potential of the generalized Hubbard model:
at $n<4/3$ increasing correlated hopping $T_1$ (or $T_2$) leads to the
increase of chemical potential, and for $n>4/3$ chemical potential
decreases with increasing correlated hoppings $T_1,\ T_2$. This can be
explained by the following arguments. For the electron concentration
$n\to 1^{+}$ the chemical potential is located at the bottom of the upper
Hubbard band. When the electron concentration is $n\to 2^{-}$, the chemical
potential position corresponds to the top of the upper Hubbard band. 
For the electron concentration $n\to 1^{+}$ chemical potential of the 
generalized Hubbard model is larger than chemical potential of the Hubbard
model, and vise versa for the electron concentration $n\to 2^{-}$,
as a result of the fact that width of the upper band of the Hubbard 
model is larger than width of the upper band of the model with electron-hole 
asymmetry. Thus, the value $n_0$ of electron concentration must exist
satisfying the requirements: at this value the chemical potentials of the
Hubbard and generalized Hubbard models are equal (in this case the chemical
potential is located at the centre of the upper Hubbard band, i.e. $\mu =U$),
for electron concentration $n<n_0$ increasing correlated hoppings $T_1,\ T_2$ 
leads to the increase of chemical potential, and for $n>n_0$ chemical 
potential decreases with increasing correlated hoppings $T_1,\ T_2$.

Similarly, we can explain the dependence of chemical potential of the 
generalized Hubbard model on the correlated hopping $T_1$ for electron 
concentration $n<1$. But in this case, with contrast to the previous 
situation, the dependence of chemical potential on the correlated hopping 
$T_1$ is weak (besides, for the doped Mott-Hubbard insulators with $n<1$
chemical potential has no dependence on the correlated hopping $T_2$).
It should be stressed that the found values $n_0=2/3,\ 4/3$ are the electron 
concentrations corresponding to the change of conductivity type of the 
system~\cite{1_6}.
 
From Figs.~\ref{mu_tau1}, \ref{mu_tau2} one can see that for electron
concentrations $n>1$ the chemical potential of the generalized Hubbard
model has the linear dependence on correlated hopping parameters, and this
dependence on $\tau_2$ is stronger versus $\tau_1$. For the values of
correlated hopping parameters $\tau_1=0,\ \tau_2=0.5$ the chemical potential
of the generalized Hubbard model at $n>1$ becomes equal to the chemical
potential of the Hubbard model in the atomic limit $\mu =U$ 
(though $t_0\neq 0$); in this case chemical potential of the generalized
Hubbard model does not depend on electron concentration. Notice that for
these or close values of correlated hopping parameters within the model
under consideration the following situation, which is analogous one observed
in heavy fermion substances and compounds with mixed valency, can occur.
Near the Fermi level which lies within the lower Hubbard band, the level or
very narrow upper Hubbard band (being analogous to the $f$-level in heavy fermion 
substances and compounds with mixed valency) can arise in conduction electron
subsystem, as a result of electron-electron interactions of correlated 
hopping type. In particular, this can lead to the very strong enhancement of 
effective mass of carriers~\cite{1_26,dhs}.

It should be noted that using more realistic expressions for the density of
states does not change essentially the behaviour character of the obtained
dependences of chemical potential on the model parameters. For example,
in the case of semielliptic density of states:
\begin{eqnarray} 
{1\over N}\sum_{\bf k}\delta(E-t({\bf k}))={2\over \pi w}
\sqrt{1-\left({E\over w}\right)^2}
\end{eqnarray}
we obtain the following equation for chemical potential of the generalized 
Hubbard model with correlated hopping:
\begin{eqnarray} 
&&{2-n\over 2\pi}\arcsin \left({2-2n+n^2\over 2-n}{\mu\over w(n)}\right)
\\
&&+{2-2n+n^2\over \pi}{\mu\over w(n)}\sqrt{1-\left({2-2n+n^2\over 2-n}\right)^2
{\mu^2\over w^2(n)}} ={3n\over 4}-{1\over 2} \qquad (n<1),
\nonumber
\\
&&{n\over 2\pi}\arcsin {(\mu-U)n\over (2-2n+n^2)\tilde{w}(n)}
\\
&&+{(\mu -U)n^2\over 2\pi(2-2n+n^2)\tilde{w}(n)}
\sqrt{1-\left({(\mu -U)n\over (2-2n+n^2)\tilde{w}(n)}\right)^2} 
={3n\over 4}-1 \qquad (n>1).
\nonumber
\end{eqnarray}
Analysis of these equations shows that the obtained behaviour peculiarities 
of chemical potential for the case of the rectangular density of states 
persist. Moreover, a choice of the semielliptic density of states does not
change even the values $n_0=2/3,\ 4/3$.

\subsection{Temperature dependence of chemical potential}

Find the dependence of chemical potential of the generalized Hubbard model
with correlated hopping upon temperature. Using the rectangular density
of states, we rewrite Eq.~(\ref{ch_p_eq}) for chemical potential in the
form:
\begin{eqnarray} 
{n\over 4\tilde{w}(n)}\int\limits_{-\tilde{w}(n)}^{\tilde{w}(n)}{d\epsilon\over 
\exp{E_2(\epsilon)\over \theta}+1} - 
{1-n/2\over 2w(n)}\int\limits_{-w(n)}^{w(n)}{d\epsilon\over 
\exp{-E_1(\epsilon)\over \theta}+1} =n-1. 
\end{eqnarray}
After integration for the case of the doped Mott-Hubbard insulator we obtain
the following equation for chemical potential:
\begin{eqnarray} \label{muT<}
&&{1\over 4\tilde{w}(n)}\ln{1+\exp{-\mu+U-n\tilde{w}(n)\over \theta}\over 
1+\exp{-\mu+U+n\tilde{w}(n)\over \theta}}=
{n/2-1\over 2\beta w(n)}\ln{1+\exp{\mu+\beta w(n)\over \theta}\over 
1+\exp{\mu-\beta w(n)\over \theta}} \quad (n<1),
\\
&&{1\over 4\tilde{\beta}\tilde{w}(n)} \label{muT>}
\ln{1+\exp{-\mu+U-\tilde{\beta}\tilde{w}(n)\over \theta}\over 
1+\exp{-\mu+U+\tilde{\beta}\tilde{w}(n)\over \theta}}=
-{1\over 4w(n)}\ln{1+\exp{\mu+ (2-n)w(n)\over \theta}\over 
1+\exp{\mu- (2-n)w(n)\over \theta}} \enspace (n>1),
\end{eqnarray}
with the coefficients $\beta ={2-2n+n^2\over 2-n},\ 
\tilde{\beta}={2-2n+n^2\over n}$. 

\begin{figure}[hpb] 
\begin{minipage}[b]{75mm}
\epsfxsize=75mm
\epsfysize=65mm
\epsfclipon
\epsffile{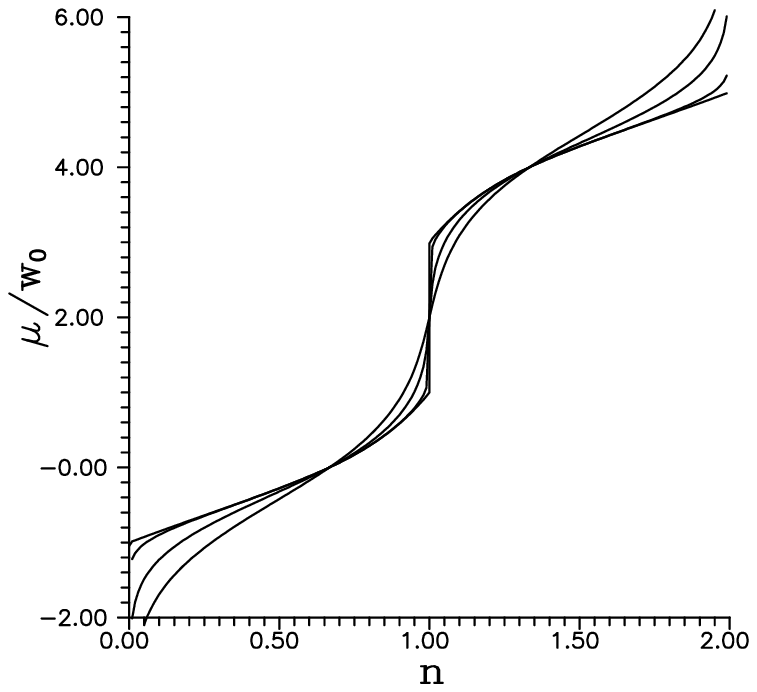}
\renewcommand{\baselinestretch}{1}
\caption{Evolution of the concentration dependence of chemical potential of
the Hubbard model with temperature for $U/w_0=4$: 
$\theta/w_0=k_BT/w_0=0,\ 0.1,\ 0.3,\ 0.5$ respectively.} \label{Tmu_n}
\end{minipage}
\hfill
\begin{minipage}[b]{75mm}
\epsfxsize=75mm
\epsfysize=65mm
\epsfclipon
\epsffile{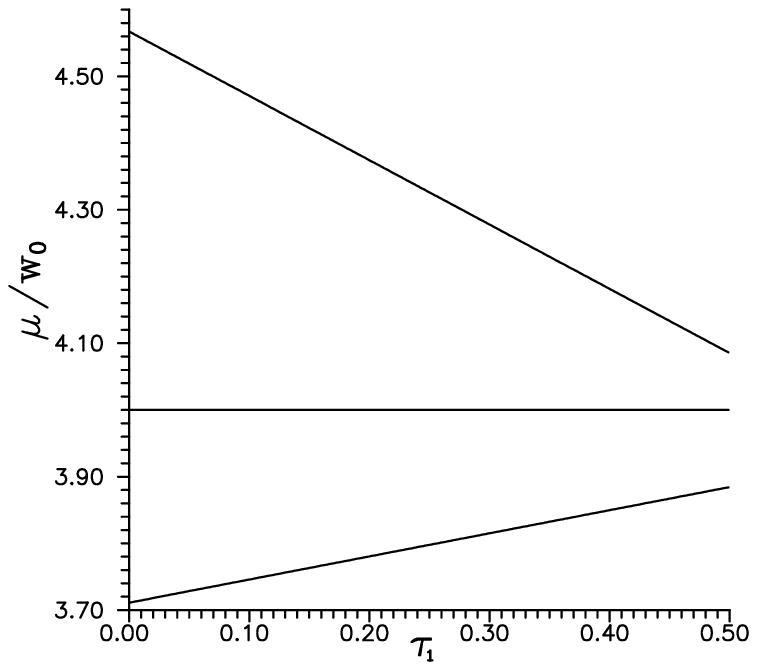}
\renewcommand{\baselinestretch}{1}
\caption{Chemical potential of the generalized Hubbard model as a function
of the correlated hopping parameter $\tau_1=T_1/|t_0|$ for $U/w_0=4$:
the lower curve corresponds to $n=1.7$, the middle -- $n=4/3$, 
the upper -- $n=1.2$.} \label{mu_tau1}
\end{minipage}
\vfill	
\begin{minipage}[t]{75mm}
\epsfxsize=75mm
\epsfysize=65mm
\epsfclipon
\epsffile{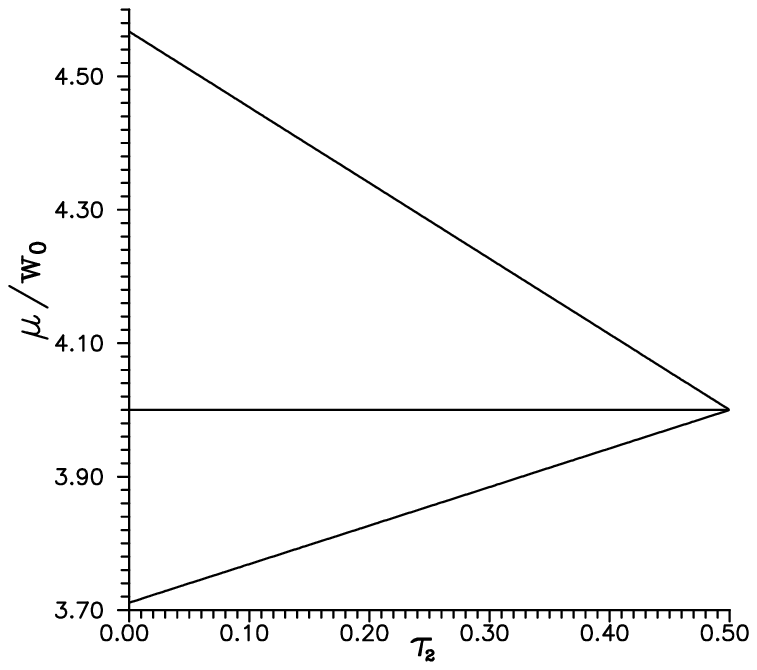}
\renewcommand{\baselinestretch}{1}
\caption{Chemical potential of the generalized Hubbard model as a function
of the correlated hopping parameter $\tau_2=T_2/|t_0|$ for $U/w_0=4$:
the lower curve corresponds to $n=1.7$, the middle -- $n=4/3$, 
the upper -- $n=1.2$.} \label{mu_tau2}
\end{minipage}
\hfill
\begin{minipage}[t]{75mm}
\epsfxsize=75mm
\epsfysize=65mm
\epsfclipon
\epsffile{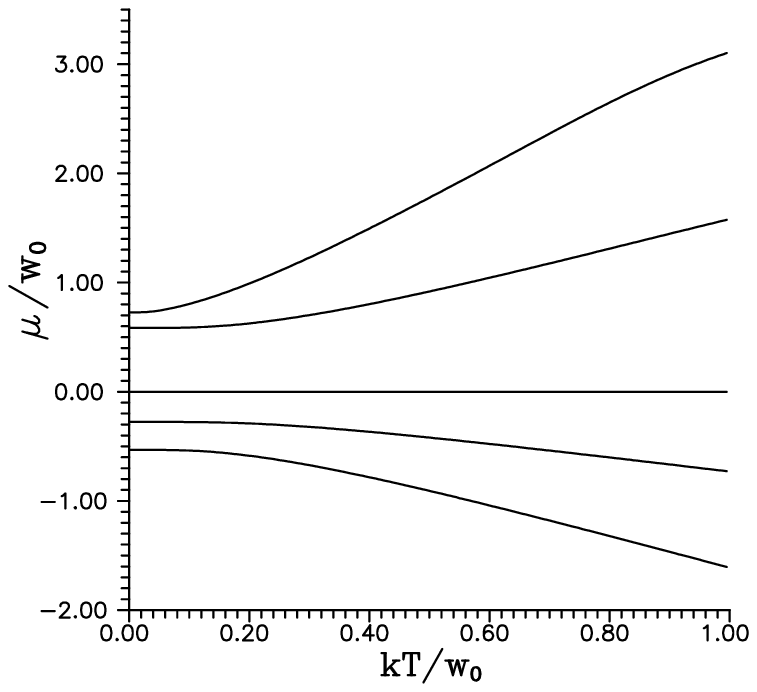}
\renewcommand{\baselinestretch}{1}
\caption{Temperature dependence of the chemical potential of the Hubbard 
model with $U/w_0=8$ for different values of electron concentration:
$n=0.98,\ 0.9,\ 2/3,\ 0.5,\ 0.3$ (from top to bottom).} \label{nmu_T}
\end{minipage}
\end{figure}

On the base of expressions~(\ref{muT<}) and 
(\ref{muT>}) we obtain the dependence of chemical potential on electron
concentration at some values of temperature (Fig.~\ref{Tmu_n}), and
$\mu (T)$-dependence for some values of electron concentration 
(Fig.~\ref{nmu_T}). Here we limit ourselves to a consideration of the Hubbard
model case, because taking into account correlated hopping of electrons 
at $T\neq 0$ does not lead to qualitatively new distinctions of the chemical
potential behaviour from the mentioned one at $T=0$.

From Fig.~\ref{Tmu_n} one can see that at temperature $T>0$
the discontinuous jump in chemical potential, obtained in the previous 
subsection, becomes more smooth, the dependence of chemical potential 
on electron concentration becomes a S-shaped curve with flex point at 
the electron concentration $n=1$. 

Figs.~\ref{Tmu_n}, \ref{nmu_T} show that for the electron concentration 
$n<n_0$ chemical potential of the doped Mott-Hubbard insulator decreases 
with the increase of temperature, and for $n>n_0$ the increase of 
temperature leads to the increase of chemical potential; at $n=n_0$ the 
chemical potential does not depend on temperature. These peculiarities of 
the temperature-dependence of chemical potential can be explained by the 
following reasons. Let us consider the case $n<1$. For small values of 
electron concentration $n<n_0$ the Fermi energy lies below the centre of the 
lower Hubbard band, thus with the increase of temperature electrons become to 
occupy the above-lied empty energy levels, this leads to decreasing
chemical potential (an analogue of the case of free electrons). For the
electron concentration $n=n_0$ the Fermi energy is located at the centre
of the lower Hubbard band, as the temperature increases the above-lied empty 
energy levels within the lower Hubbard band become to be occupied, but
the chemical potential does not change its value as a result of symmetry
about the centre of the band. For enough high temperatures and values of
the intra-atomic Coulomb repulsion $U\geq 2w_0$ some electrons are thermally 
activated into the upper Hubbard band, therefore the chemical potential value
decreases with increasing temperature at $n=n_0$. For values of electron 
concentration $n>n_0$ the number of empty energy levels within the lower
Hubbard band is small because the Fermi energy in this case lies above the
centre of the lower band, consequently with the increase of temperature
chemical potential increases owing to occupancy of these levels and a presence
of the empty upper Hubbard band. Similarly, the influence of temperature and
electron concentration on chemical potential of the Hubbard model for the 
values of electron concentration $n>1$ can be explained.

Note that one can see an analogy of the temperature dependence
of chemical potential of the model under consideration with the same
dependence in doped semiconductors. Really, for electron concentration
$n<n_0$ within the model the system has $n$-type of conductivity~\cite{1_6}
and chemical potential decreases with increasing temperature as in $n$-type
semiconductors. For electron concentration $n>n_0$ the system has $p$-type
of conductivity, correspondingly chemical potential increases with the
increase of temperature, as it is characteristically for $p$-type 
semiconductors.

\section{Conclusions}

In the present paper chemical potential of the generalized Hubbard model with 
correlated hopping has been studied by means of a generalized mean-field
approximation. The peculiarity of the model in comparison with similar 
generalized Hubbard models is the concentration dependence of hopping 
integrals caused by taking into consideration the matrix elements of 
electron-electron interaction which describe inter-site hopping of electrons
(correlated hopping). 

The chemical potential as a function of the model energy parameters 
($U,\ t(n),\ \tilde{t}(n),\ t'(n)$), electron concentration and temperature 
has been found. At half-filling, in contrast with the Hubbard model 
where $\mu =U/2$, the dependences of chemical potential on energy parameters 
of the model are different in metallic and insulating phases leading to a kink 
at the point of metal-insulator transition at zero temperature; with the 
increase of temperature the kink in the chemical potential curve disappears.
With the increase of the correlated hopping $T_2$ (i.e. with decreasing 
$\tilde{w}$) chemical potential moves towards the upper Hubbard band.

Chemical potential of the generalized Hubbard model with correlated hoping
is temperature-dependent even in the case of half-filled band (in contrast 
to the Hubbard model). With the decrease of temperature chemical 
potential rapidly increases, this dependence is different for the different
values of correlated hopping parameters $\tau_1,\ \tau_2$. In high temperature
region chemical potential of the asymmetric Hubbard model tends to $U/2$
as temperature increases.

In strong coupling regime we have found that correlated hopping changes 
essentially the location of chemical potential of the generalized Hubbard
model: for the electron concentration $n<n_0$ 
increasing correlated hopping parameters leads to the increase of 
chemical potential, and for the case $n>n_0$ chemical potential decreases 
with increasing correlated hopping parameters. 
Similarly, for the electron concentration $n<n_0$ chemical potential 
decreases with the increase of temperature, and for $n>n_0$ the increase of 
temperature leads to the increase of chemical potential; at the value of 
electron concentration $n=n_0$ the chemical potential does not depend on 
correlated hopping and temperature. The found values $n_0=2/3,\ 4/3$ 
correspond to the electron concentration when conductivity type of the 
system changes; chemical potential of the systems with these electron 
concentrations is located at the centre of the lower and upper Hubbard band
respectively.

At zero temperature we have found the jump in chemical potential at 
half-filling in going from electron to hole doping, this discontinuity in 
chemical potential is strictly equal to the energy gap width 
$\Delta E=U-w-\tilde{w}$.
If the temperature is greater than zero then the discontinuous jump of 
chemical potential becomes more smooth, the dependence of chemical potential 
on electron concentration becomes a S-shaped curve with flex point at 
the electron concentration $n=1$.

\newpage
\newpage

\end{document}